\documentstyle[epsf,aps,prl,multicol]{revtex}

\draft

\begin{document}

\title{Neutron scattering and molecular correlations in a
supercooled liquid}

\author{Christoph Theis and Rolf Schilling}

\address{Institut f\"ur Physik, Johannes Gutenberg--Universit\"at,
Staudingerweg 7, D--55099 Mainz, Germany.}

\maketitle

\begin{abstract}

We show that the intermediate scattering function $S_n(q,t)$ for neutron
scattering (ns) can be expanded naturely with respect to a set of molecular
correlation functions that give a complete description of the translational
and orientational two--point correlations in the liquid. The general
properties of this expansion are discussed with special focus on the
$q$--dependence and hints for a (partial) determination of the molecular
correlation functions from neutron scattering results are given. The resulting
representation of the static structure factor $S_n(q)$ is studied in detail
for a model system using data from a molecular dynamics simulation of a
supercooled liquid of rigid diatomic molecules. 
The comparison between the
exact result for $S_n(q)$ and different approximations that result from a 
truncation of the series representation demonstrates its good convergence for
the given model system. On the other hand it shows explicitly that the
coupling between translational (TDOF) and orientational degrees of freedom 
(ODOF) of each molecule and rotational motion of different molecules can not
be neglected in the supercooled regime.
Further we report the existence of a prepeak in the ns--static structure
factor of the examined fragile glassformer, demonstrating that prepeaks can
occur even in the most simple molecular liquids. Besides examining the
dependence of the prepeak on the scattering length and the temperature we
use the expansion of $S_n(q)$ into molecular correlation functions to point
out intermediate range orientational order as its principle origin.
\end{abstract}


\begin{multicols}{2}

\section{Introduction}

Neutron scattering is one of the most important tools to determine the
structure and dynamics of condensed matter. One of its major advantages is
that the neutron as an electrically neutral probe is not influenced by the
electron cloud of the target atoms but interacts only with the nucleus.
Consequently neutron scattering can be directly interpreted in terms of the
atomic structure and motion. However, if one wants to examine a molecular
system not only with respect to the constituent atoms but in terms
of the molecular units some care has to be taken. Since
the superposition of scattering from atomic sites is in general not
equivalent to scattering from the molecular center of mass, neutron
scattering from molecular liquids is sensitive to orientational as well as
translational correlations in the system.

This was realized already many years ago
\cite{sachs41,zemach56,krieger57,sears67,powles73} 
and led to different formulations
of neutron scattering by molecules. For reasons of simplicity and also
because the center of mass correlations were of principle interest those
approaches tried to "correct" for the effect of orientational degrees of
freedom. In the work of Sears \cite{sears67} this is done using the
so called {\em weak hindering approximation}, i.e. the
assumptions of statistical independence of {\em i)} rotational and
translational motion of any molecule and 
{\em ii)} rotational motion of any two
different molecules.

In recent years the mode coupling theory of the glass transition
(see e.g. \cite{gotze91,gotze92,schilling94}) stimulated a lot 
of experiments on supercooled
liquids. Among other techniques also neutron scattering was used
(see e.g. \cite{richter89} and for more recent work
\cite{tolle97,wuttke95}) to study the static and dynamic properties of
glassforming substances most of which are of molecular nature, like e.g.
glycerol, salol or orthoterphenyl. In the situation of the dense packed
molecular liquid where the motion becomes more and more cooperative the
assumptions of Sears are not reasonable anymore as was shown by Chen et al.
\cite{chen97} for incoherent neutron scattering in simulated supercooled
water. Below, in the study of a liquid of diatomic rigid molecules we will
come to the same conclusion for coherent scattering. 
This raises the question how neutron
scattering data can be analyzed in terms of the molecules in this case.

With increasing computer power the simulations of supercooled liquids
trend to deal with more realistic 
model systems \cite{kob95} like SPC/E water
\cite{sciortino96} or orthoterphenyl \cite{lewis94}. Thus it has become
feasible to examine in detail also the orientational degrees of freedom
in the strongly supercooled regime (see e.g. \cite{kammerer,fabbian98}). 
In experiments it is up to now only possible to measure orientational 
correlators for $q=0$, i.e. one does not get information about the 
spatial arrangement of orientationally correlated molecules. On the
other hand this information would be valuable to understand the
interaction between translational and orientational degrees of
freedom. Therefore it seems sensible to examine in which way
orientational correlations enter into the neutron scattering function.

A feature of special interest in supercooled liquids and glasses is the
appearance of a prepeak in the static ns--structure factor at a $q$--value
that corresponds to distances larger than the average nearest neighbour
distance. As a sign of intermediate range order prepeaks have been studied
in a variety of systems (see e.g.
\cite{wilson94,elliott91,salmon94} and references therein). 
Mostly they have been
attributed to the network structure of strong glasses or charge effects in
ionic glassformers. The behaviour of prepeaks in molecular systems seems to 
be quite different as was pointed out recently by Morineau et al. 
\cite{morineau98} in their study of the temperature-- and pressure--dependence
for m--toluidine and m--fluoroaniline which will be discussed in 
section \ref{prepeak}.

We will present results for the ns--static structure factor of a liquid of
diatomic molecules, i.e. one of the most simple systems in
which effects of the molecular nature can be studied. 
We will discuss the dependence of the prepeak 
on temperature and scattering
length and make comparison to findings for other systems. Using the
representation of $S_n(q)$ in terms of the molecular correlation functions
we will also give strong evidence that the existence of the peak is
closely connected to orientational degrees of freedom.

The paper is organized as follows. In Section \ref{theory} we will
summarize shortly the approach of Sears 
to the intermediate scattering function
$S_n(q,t)$ for molecular systems. We will show that dropping the weak
hindering approximation leads naturely to a description in terms of
molecular correlation functions. Some aspects of the representation for
arbitrary molecules are discussed and hints for an application to 
experiments are given. Section \ref{results} presents the
results from the molecular dynamics (MD) simulation of a system of diatomic
molecules. It is divided into three subsection concerning respectively the
molecular correlation functions, the ns-static structure factor and a
detailed discussion of the prepeak. The final section gives some 
conclusions.

\section{Neutron Scattering Function}
\label{theory}

We consider a set of $N$ rigid molecules of identical geometry each
consisting of $s$ atoms. The assumption of rigidity is one that is commonly
used as well in the theoretical analysis as in computer simulation of the
liquid state. It is justified on the ground that at the temperature of the
liquid only the lowest vibrational states are populated. The starting point
of an analysis of neutron scattering from molecules is the following
site--site--representation of the intermediate scattering function ( see
e.g. \cite{sears67,hansen76})

\begin{eqnarray}
S_n(q,t) &=& \frac{1}{Ns} \sum_{j,j'=1}^N \sum_{\nu,\nu'=1}^s \left(
a_{coh}^{j\nu} a_{coh}^{j'\nu'} + (a_{inc}^{j\nu})^2 \delta_{j j'}
\delta_{\nu \nu'} \right) \; \cdot \nonumber \\
& & \;\; \cdot \;\; \langle e^{i \vec{q} \cdot (\vec{R}_{j\nu}(t) -
\vec{R}_{j'\nu'})} \rangle
\label{eq:site-site}
\end{eqnarray}
\noindent
where $a_{coh}^{j\nu}$ ($a_{inc}^{j\nu}$) is the coherent (incoherent)
scattering length and $\vec{R}_{j\nu}(t)$ the position of atom $\nu$ in
molecule $j$ at time $t$. The brackets $\langle ... \rangle$ denote the
canonical averaging over initial conditions. Unlike the geometry which is
supposed to be identical for all molecules the scattering length are
allowed to differ from molecule to molecule. This assumption is quite
realistic since the chemical structure is independent of the isotopic
composition whereas the scattering lengths are.

Appealing as this site--site--description is, it is not very favourable
from a molecular view. The rigid molecules are most suitably described by
the center of mass coordinates $\vec{R}_j(t)$ and the Euler angles
$\Omega_j(t) = (\phi_j(t),\theta_j(t),\chi_j(t))$ giving the orientation of
the body fixed system relative to the laboratory frame.

To make a connection between these two different views 
Sears \cite{sears67} introduced center of mass and
relative coordinates by $\vec{R}_{j\nu} = \vec{R}_j + \vec{r}_{j\nu}$ and
used the Rayleigh expansion of the plane waves

\begin{eqnarray}
e^{i \vec{q} \cdot \vec{r}_{j\nu}(t)} &=& 
\sum_{ln} [4\pi (2l+1)]^\frac{1}{2}
i^l j_l(q r_\nu) \; \cdot \nonumber \\
& & \; \cdot \; Y_l^n(\vartheta_\nu,\varphi_\nu) 
{\cal D}^l_{0n}(\Omega_j(t))
\label{eq:rayleigh}
\end{eqnarray}
\noindent
where $(r_\nu,\vartheta_\nu,\varphi_\nu)$ are the 
polar coordinates of atom $\nu$
with respect to the body fixed frame with origin at the center of mass of
molecule $j$. Without loss of generality the $z$--axis of the laboratory
frame has been chosen to point in direction of $\vec{q}$. The special
functions appearing in eq.(\ref{eq:rayleigh}) are the spherical Bessel
functions $j_l$, the spherical harmonics $Y_l^m$ and the Wigner functions
${\cal D}^l_{mn}$ \cite{abramowitz72,gray84}. In the summation of
eq.(\ref{eq:rayleigh}) $l$ runs over 
$0,1,2,...$ and $n$ is restricted by $-l
\le n \le l$.

Following this strategy and {\em not} doing any factorization of
translational and rotational motion or of rotational motion of different
molecules one is naturely led to consider correlation functions of the
following kind

\begin{eqnarray}
& & S_{ln,l'n'}^m(q,t) = i^{l'-l} [(2l+1)(2l'+1)]^\frac{1}{2} \frac{1}{N}
\sum_{j j'} \; \cdot \nonumber \\
& & \quad \cdot \; \langle e^{-i q (R_j^z(t)-R_{j'}^z)} {\cal
D}_{mn}^l(\Omega_j(t)) {\cal D}_{mn'}^{l' \ast}(\Omega_{j'}) \rangle,
\label{eq:smolqt}
\end{eqnarray}
\noindent
which equal $\frac{1}{N} \langle \rho_{lmn}^\ast(q,t) \; \rho_{l'mn'}(q,0)
\rangle$, the correlation function of the tensorial one particle
density mode $\rho_{lmn}(q,t)$ \cite{schilling97}.
As usual it is possible to separate them into self ($j=j'$) and distinct
terms ($j \not= j'$) giving

\begin{equation}
S_{ln,l'n'}^m(q,t) = S_{ln,l'n'}^{(s) m}(q,t) + S_{ln,l'n'}^{(d) m}(q,t).
\label{eq:self+dist}
\end{equation}

This set of correlation functions is the generalization to arbitrary
molecules of the correlators introduced in the mode coupling theory
for a single molecule in an atomic liquid \cite{franosch97} and for
liquids of linear molecules \cite{schilling97} and
has been used recently to study molecular
correlations in supercooled water \cite{fabbianpp} and diatomic
molecules \cite{kammerer}.
Similar correlation functions have also been used earlier in the study of
molecular liquids \cite{hansen76,gray84} and go back to the work
on statistical mechanics of nonspherical molecules by Steele
\cite{steele63}.

On first sight one might be scared by the large number of indices and not
willing to consider an infinite set of correlation functions. Therefore we
want to stress that due to the tensorial nature of the
quantities (\ref{eq:smolqt}) the symmetry of the molecule may reduce the
number of correlators. Secondly one can expect that reasonable results can
be achieved already if one considers only the finite number of correlators
with $l,l' \le l_{co}$ for small $l_{co}$. For a system of diatomic
molecules we will see in section \ref{results} that $l_{co}=4$ turns
out to be sufficient to describe $S_n(q)$ in the $q$--range of
interest. Depending on the size and form of the molecules it might be
necessary to consider also larger values for $l_{co}$. This becomes
clear if one looks at the angular dependence of the Wigner functions since
extending the range of $l$ corresponds mainly to improving the angular
resolution. In this argument we have considered the factors ${\cal D}_{m
n}^l(\Omega)$ in eq.(\ref{eq:smolqt}) 
as "weights" which select molecules within a certain range of
orientations for the canonical averaging. While for instance ${\cal
D}_{0 0}^0(\Omega) \equiv 1$ gives equal weight to all orientations,
${\cal D}_{0 0}^1(\Omega) \propto Y_1^0(\theta,\phi)$ puts emphasis on
molecules with $\theta \approx 0$ or $\theta \approx \pi$.
This may elucidate a bit more the meaning of the $q$--dependent 
orientational correlators.

Having introduced the suitable correlation functions
eqs.(\ref{eq:site-site})--(\ref{eq:smolqt}) can be easily combined to give

\begin{eqnarray}
S_n(q,t) &=& \sum_{l l'} \sum_{n n'} [ 
b_{ln,l'n'}^{inc}(q) S_{ln,l'n'}^{(s) 0}(q,t) + \nonumber \\
& & \; + b_{ln,l'n'}^{coh}(q) S_{ln,l'n'}^{(d) 0}(q,t) ]
\label{eq:snresult1}
\end{eqnarray}
\noindent
with the coefficients

\begin{eqnarray}
& & b_{ln,l'n'}^{inc}(q) = \frac{1}{s} \sum_{\nu \nu'} 4 \pi j_l(q r_\nu)
j_{l'}(q r_{\nu'}) \nonumber \\
& & \quad Y_l^{n \ast}(\vartheta_\nu,\varphi_\nu)
Y_{l'}^{n'}(\vartheta_{\nu'},\varphi_{\nu'}) [\overline{a_{coh}^\nu
a_{coh}^{\nu'}} + \overline{(a_{inc}^\nu)^2} \delta_{\nu \nu'} ]
\label{eq:binc} \\
& & b_{ln,l'n'}^{coh}(q) = \frac{1}{s} \sum_{\nu \nu'} 4 \pi j_l(q r_\nu)
j_{l'}(q r_{\nu'}) \nonumber \\
& & \quad Y_l^{n \ast}(\vartheta_\nu,\varphi_\nu)
Y_{l'}^{n'}(\vartheta_{\nu'},\varphi_{\nu'}) \overline{a_{coh}^\nu}
\overline{a_{coh}^{\nu'}}.
\label{eq:bcoh}
\end{eqnarray}
\noindent
Here $\overline{x} = 1/N \sum_j x^j$ denotes the average over molecules.
For the special case of a single linear molecule in an isotropic liquid
a similar result has been
given by Franosch et al. \cite{franosch97}. 

The representation eq.(\ref{eq:snresult1}) has a number of interesting
properties. Immediately obvious is that it can be Fourier transformed to
give a corresponding relation for the spectra or susceptibilities with the
same coefficients. Further we notice that only correlators with $m = 0$
enter into $S_n(q,t)$ which is a consequence of the isotropy of the fluid.
From the general expression one can also draw some conclusions about the
$q$--dependence of the ns--intermediate scattering function.
Eq.(\ref{eq:snresult1}) clearly reflects that neutron scattering from
molecular liquids is the superposition of two contributions: $i)$ the
correlations within a molecule which are expressed in the incoherent terms
involving the self--part of the molecular correlation functions and $ii)$
the correlation between molecules which enter the distinct terms. The
$q$--dependence of both contributions again has two sources. The
coefficients $b_{ln,l'n'}^{inc}(q)$ and $b_{ln,l'n'}^{coh}(q)$ which could
be termed {\em incoherent} and {\em coherent molecular form factors}
respectively are
completely determined by the molecular geometry and the scattering lengths.
Thus they can be calculated exactly if the molecular units are known. Their
$q$--dependence is given by the spherical Bessel functions $j_l(q r_\nu)$
and thus connected primarily to the distances $r_\nu$ of the atoms from the
center of mass of the molecule. 
We will discuss the $q$--dependence of the molecular form factors in
more detail in the following section. 
The more important quantities that enter
into the $q$--dependence of $S_n(q,t)$ are of course the molecular
correlation functions which give a statistical description of the
interactions and the dynamic of the system.
We will discuss their $q$--dependence for the mentioned model system
in the first subsection of the following section.

The coefficients $b_{ln,l'n'}^{coh}(q)$ and $b_{ln,l'n'}^{inc}(q)$ 
represent weighting functions that
determine at which $q$--values a given molecular correlation function
$S_{ln,l'n'}^0(q,t)$ appreciably contributes to $S_n(q,t)$. 
In turn knowledge of the molecular form factors could be used to
attribute the structure of the ns--intermediate scattering function to some
molecular correlation functions. 
To undertake such an effort one would
have to do a series of experiments in which the molecular form 
factors $b_{ln,l'n'}^{coh}(q)$ and $b_{ln,l'n'}^{inc}(q)$ 
are varied systematically.
A common technique to do so would be the use of isotopes of different 
scattering length, i.e. to combine results obtained
with mixtures of different isotopic composition. From the discussion of
eq.(\ref{eq:snresult1}) given above we 
can conclude the following limitations:
$i)$ only molecular correlation functions $S_{ln,l'n'}^0(q,t)$ with $m = 0$
are accessible , $ii)$ information about the correlator
$S_{ln,l'n'}^0(q,t)$ can only be extracted from $S_n(q,t)$ in a $q$--range
where the form factor $b_{ln,l'n'}(q)$ is different from zero and not too
small compared to the other form factors. 
Upon changing the scattering lengths one can separate the site--site
correlation functions contributing to $S_n(q,t)$. Since these are
connected by a linear relation with the functions $S_{ln,l'n'}(q)$ (as
a special case of eqs.(\ref{eq:snresult1})--(\ref{eq:bcoh}))
we have $iii)$ the number of
molecular correlation functions that can be determined (for a given
$q$) is restricted by the number of site--site correlators.

Since the molecular form factors depend also on the atomic
configuration within the molecule another idea
would be to compare results for molecules with similar geometry.
To examine the structure of a liquid of diatomic molecules one could
for example try to combine results for $F_2$,$\; Cl_2$,$\; Br_2$ and $I_2$
taking into account the shift in the average nearest neighbour
distance due to the difference in atomic size by rescaling the 
$q$-values accordingly.  
In such a way one could change systematically
the intramolecular distance $d$.
Although this would offer a way to bypass the limitations given
by $iii)$ changing the geometry will also affect the molecular
correlation functions themselves.

In spite of this limitations we think that experimental efforts in the
direction of determining molecular correlation functions from
neutron scattering are worth considering. Here one has to keep in mind that
up to now information about the orientational degrees of freedom is
restricted to the few $q=0$--correlation functions that can be measured by
dielectric response or NMR. Especially to understand the interaction
between rotational and translational degrees of freedom the $q$--dependent
molecular correlation function have to be considered.

\section{Results}
\label{results}

We will now turn to the examination of a special system. On the one
hand this will give us the opportunity to illustrate the molecular
correlation functions and the representation of the scattering
function introduced in sec. \ref{theory}. On the other hand the system
shows interesting features which are apparent in the ns--static
structure factor $S_n(q)$ and give on their own a motivation for a
detailed inspection.

We will study a simulated supercooled liquid of 500 diatomic rigid
molecules each consisting of two atoms which will be labeled A and B.
The atoms have equal masses but the head--tail--symmetry is slightly
broken by the interactions which are given by a superposition of
Lennard--Jones potentials between the atomic sites. The molecular bond
length was fixed at $d = 0.5$ in units of the Lennard--Jones radius of
the A--atoms which we will use throughout this article. Particularly,
$q$ is given in units of $2 \pi$ times the inverse of that radius.
For further
details about the molecular dynamics simulation and the potential we
refer the reader to refs.\cite{kammerer}.

In the following we will concentrate on the static properties 
of the system. This
is done because of a number of reasons: 
$i)$ To test the convergence of the series representation
(\ref{eq:snresult1}) as function of
the $l$--cutoff, it is sensible to consider the worst case. Usually
this should be given by the static case as the following argument
explains. If one assumes that the relaxation of the ODOF takes place
primarily through small angular variations and that large angular jumps
can be neglected the correlators with larger $l$, i.e. better angular
resolution, will decay faster than those with small $l$. Therefore one
concludes that for $t > 0$ terms with larger $l$ are less
important than in the static case. $ii)$ Last but not least we are
primarily interested in the $q$--dependence. Thus it will be necessary
to choose a fixed time $t$ and $t = 0$ being a natural choice.

The readers interested as well in the dynamic of the studied system
and especially experimentators who are interested in the timescales
involved shall be referred to the publications of K\"ammerer et al.
\cite{kammerer} where these aspects have been discussed for TDOF as
well as ODOF.

\subsection{Molecular Correlation Functions}
\label{correlators}

We will now present molecular correlation functions for the system of
diatomic rigid molecules. As already mentioned in section \ref{theory}
the number of independent correlators can be reduced for molecules
possessing an intrinsic symmetry. In our case we deal with linear
molecules that are invariant under any rotation around the axis
connecting atoms A and B. This axis will be chosen as the $z$--axis of
the body fixed frame of reference in the following.

Symmetry considerations similar to those carried out in ref.
\cite{fabbianpp} show that the distinct part of the static correlation
functions fulfills

\begin{equation}
S_{ln,l'n'}^{(d) m}(q) = \left\{
\begin{array}{c@{\qquad}c}
S_{l l'}^{(d) m}(q) & n=n'=0 \\
0 & \mbox{otherwise} 
\end{array}
\right.
\label{dist}
\end{equation}

Since the self part of the static correlation is given simply by
$S_{ln,l'n'}^{(s) m}(q) = \delta_{l l'} \delta_{n n'}$ all the
important structure is contained in the correlation functions with
$n=n'=0$. As pointed out in section \ref{theory} for neutron
scattering only correlators with $m=0$ are relevant. Thus in the
following we will only consider the quantities $S_{l l'}(q) \equiv
S_{l0,l'0}^0(q)$. We can further restrict ourselves to $l \le l'$ since the
correlation functions $S_{l l'}(q)$ are real and 
symmetric with respect to $l$ and $l'$
as can be easily seen from their definition (\ref{eq:smolqt}).

\begin{figure}[htb]
\centerline{
\epsfxsize=8cm
\epsfysize=6cm
\epsffile{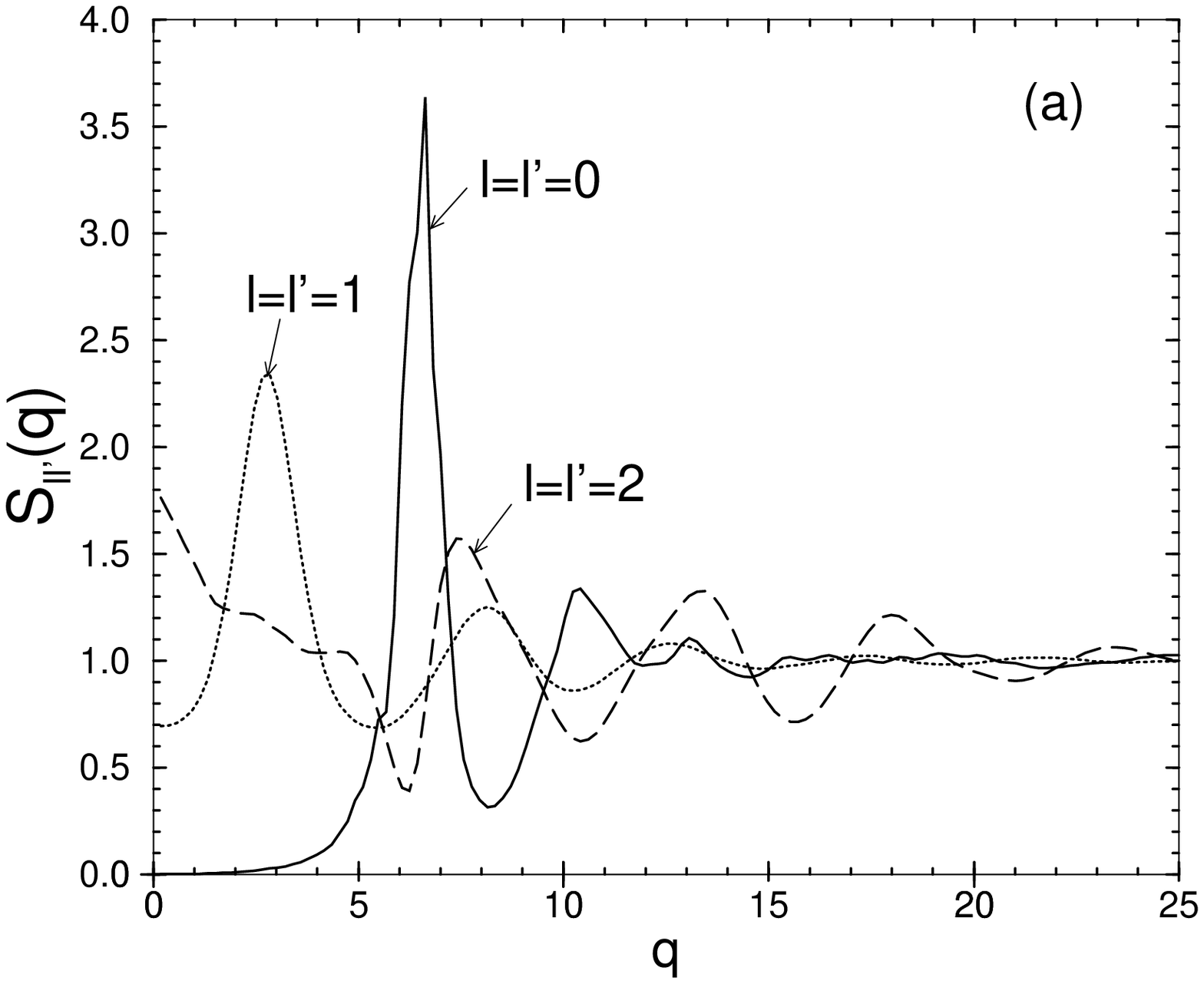} 
\hspace{0.5cm}
\epsfxsize=8cm
\epsfysize=6cm
\epsffile{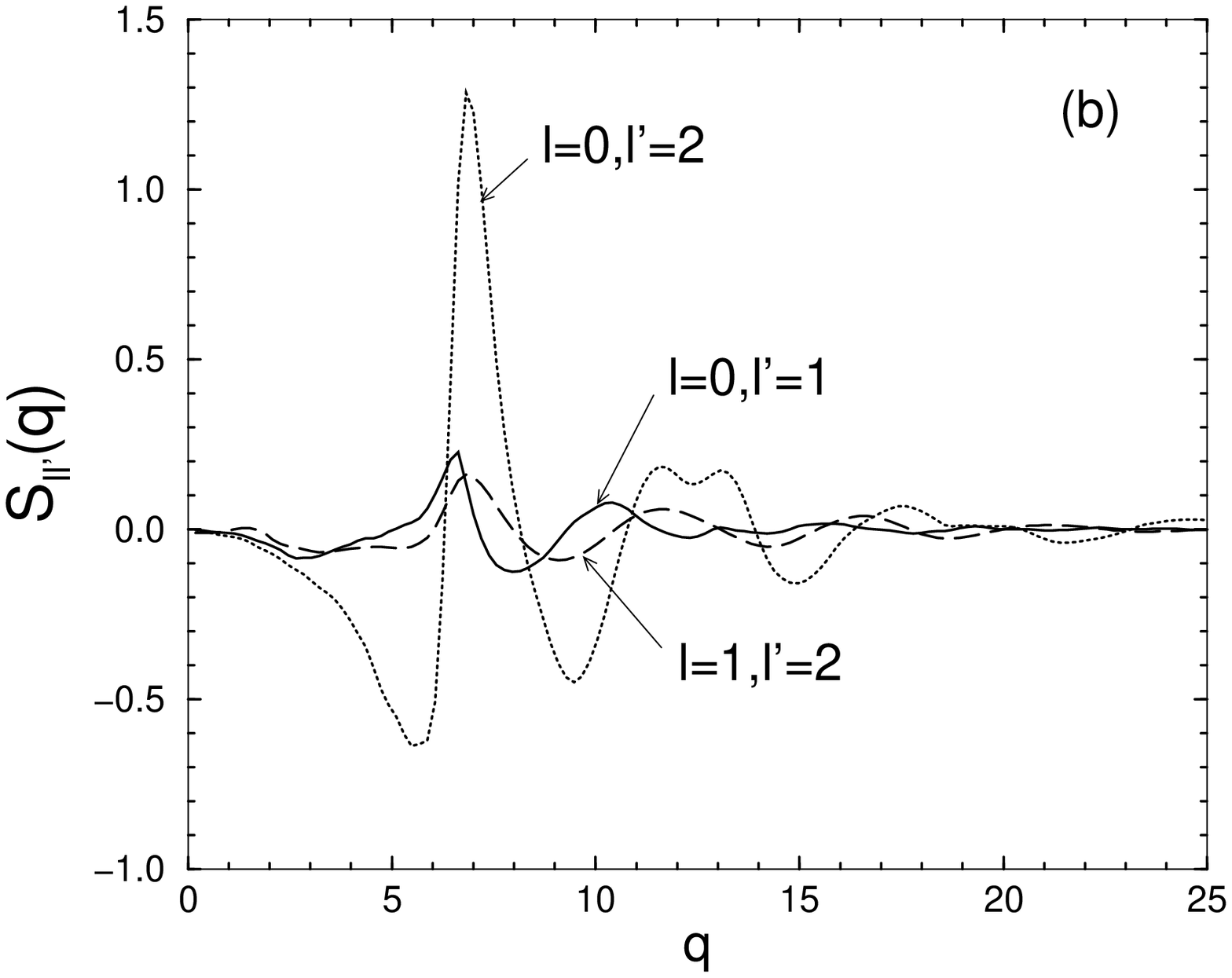}}
\centerline{
\epsfxsize=8cm
\epsfysize=6cm
\epsffile{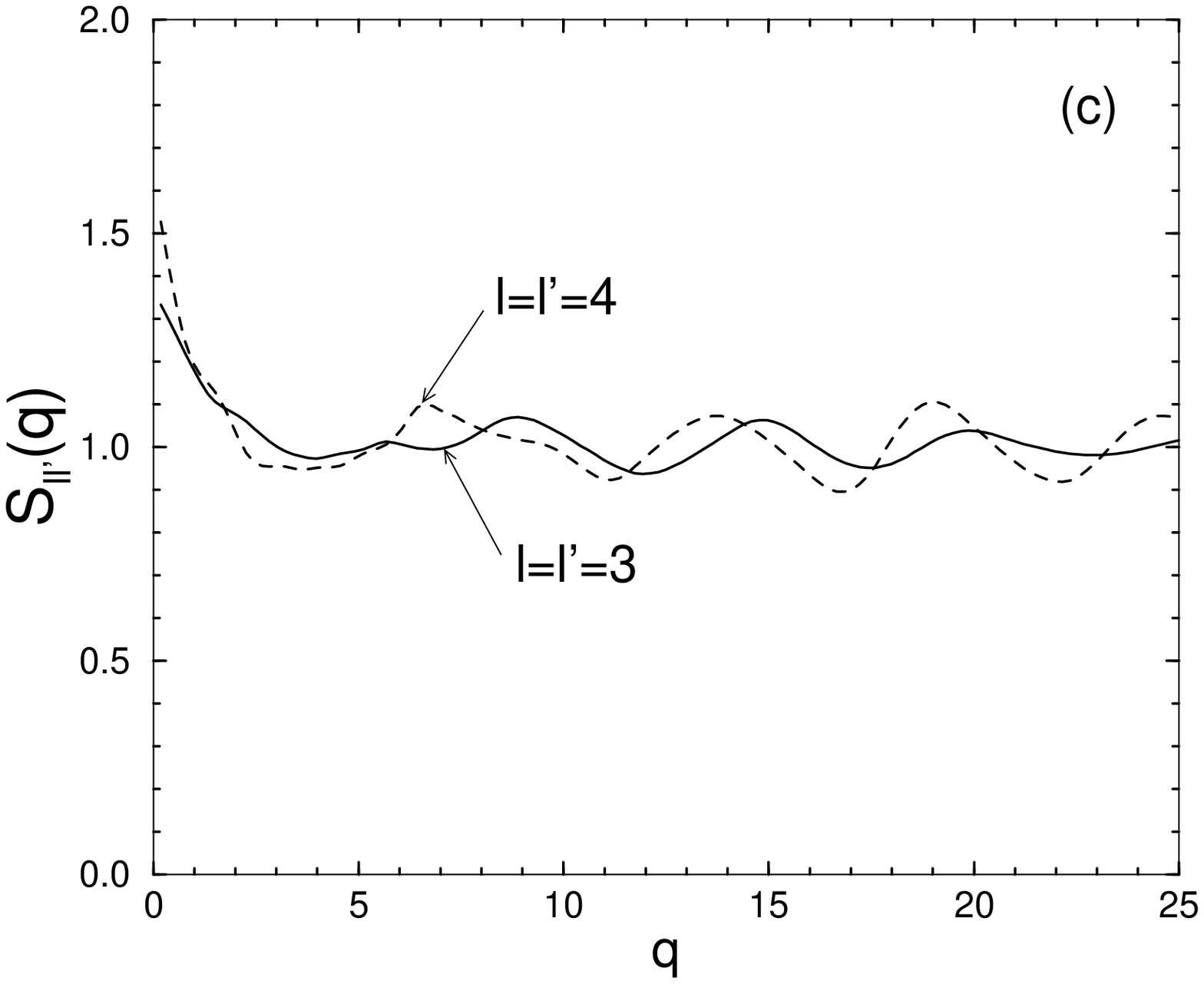}}
\centerline{
\epsfxsize=8cm
\epsfysize=6cm
\epsffile{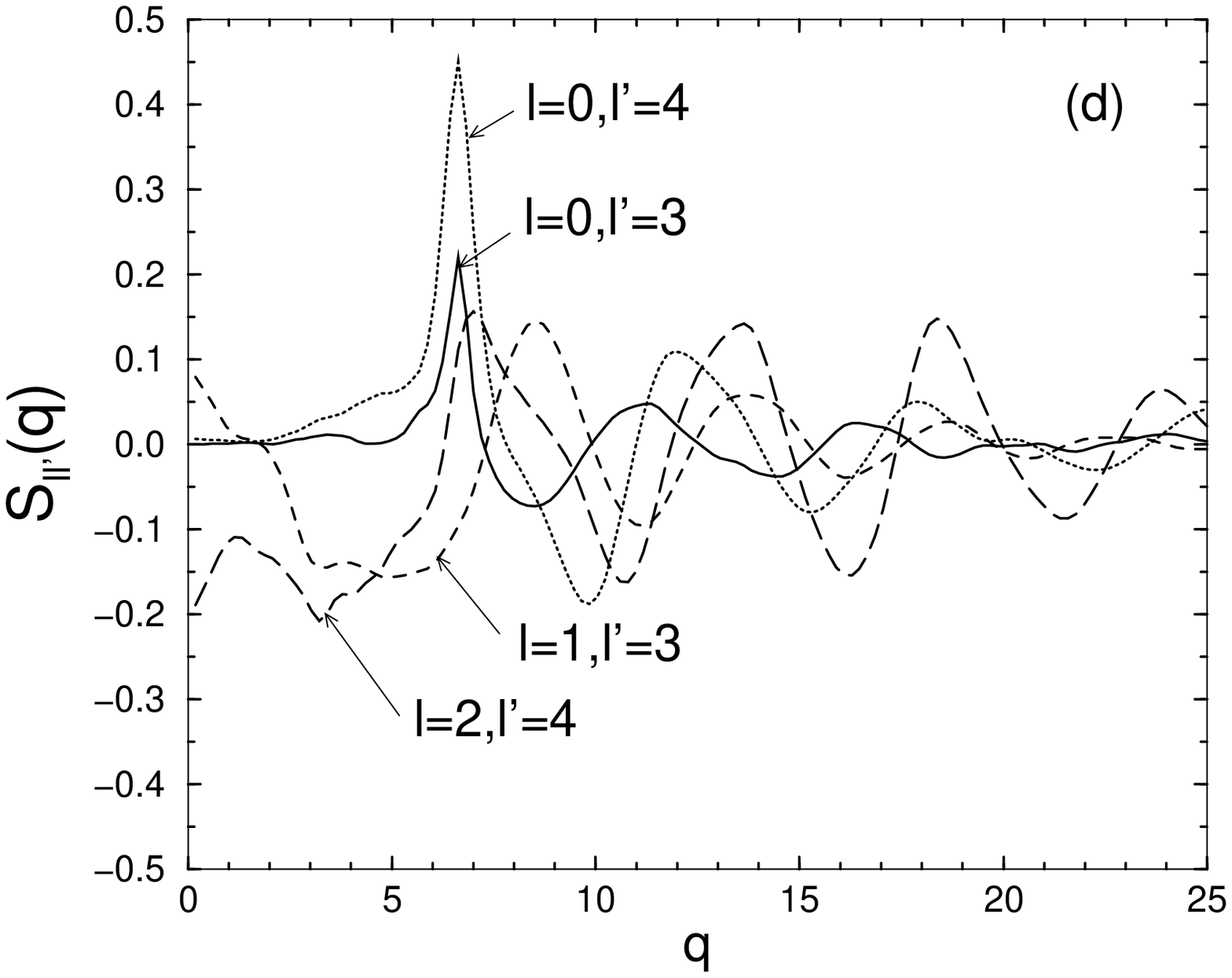}
\hspace{0.5cm}
\epsfxsize=8cm
\epsfysize=6cm
\epsffile{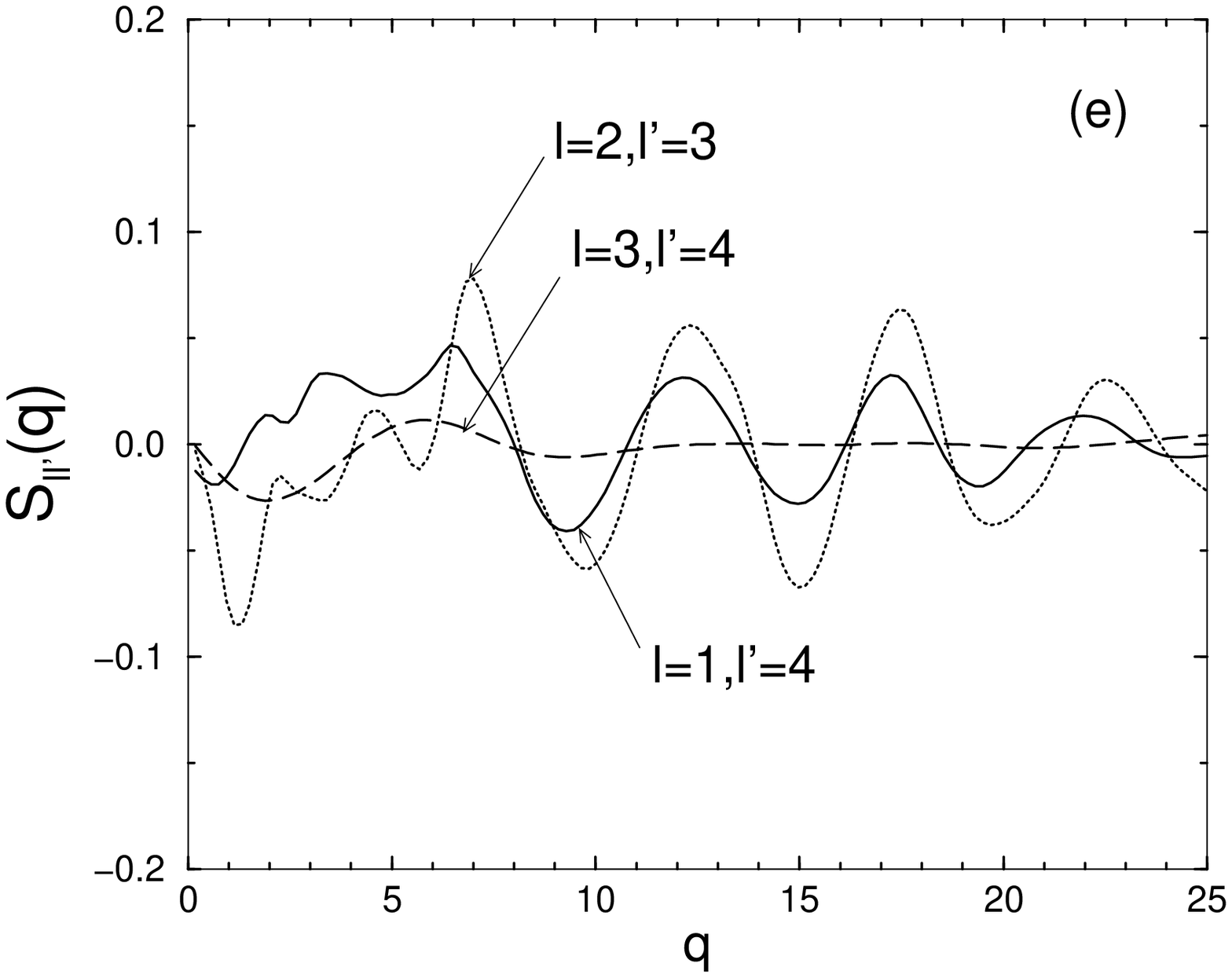}}
\caption{Molecular static structure factors $S_{l l'}(q)$ for a system 
of diatomic rigid molecules at the lowest temperature T=0.477 of the
simulation.}
\label{fig:smolq}
\end{figure}

The correlation functions up to $l,l' \le l_{co} = 2$ have
already been presented by K\"ammerer et al. \cite{kammerer}.
We want to take the opportunity to point out a minor error in this
publication: In order to get the correct data which is in accordance
with the chosen conventions (ours and theirs) the offdiagonal terms
given in Fig.3 of the last reference of \cite{kammerer} have to be
multiplied by -1. The correct graphs for the $m=0$--terms are given in
our Fig.\ref{fig:smolq} (b).

In addition we have determined all correlators up to $l,l' \le l_{co}
= 4$. Figures \ref{fig:smolq} (c)-(e) show the
$m=0$--terms which are relevant for neutron scattering. In the
discussion given below we will only refer to this data although the
properties of the $m \not= 0$--terms (not shown here) are quite
similar.

All the molecular correlation functions have been evaluated at
$T=0.477$ (in units of the Lennard--Jones energy $\epsilon_{AA}$ of
the A--particles) which is the lowest temperature considered in the
simulation. This is a temperature located in the supercooled regime
very close to the critical temperature $T_c = 0.475$ of the mode
coupling theory \cite{kammerer}.

We will now give a discussion of the
molecular correlators $S_{l l'}(q)$ by comparing their $q$--dependence.
Thereby we want to point out the similarities and differences between
the $q$--dependent orientational correlators ($l$ and/or $l'$ different 
from zero) and the better known center of mass correlations 
($l = l' = 0$). We also mention some features for which a general 
relation between the $q$--dependence and the values of $l$ and $l'$ 
seems to exist.
Thereby we will substantially enlarge the discussion given
by K\"ammerer et al. which focussed mainly on the $m$--dependence and
the effects of the approximate head--tail symmetry of the molecule.

The most prominent peak is displayed by the center of mass correlator
$S_{0 0}(q)$. Located at a $q$--value of 
$q_{max} \approx 6.6$ it represents
the first order of the nearest neighbour peak. Such a peak can also
be found at almost exactly the same position for all correlation functions
$S_{0 l'}(q)$ with $l = 0$. Also for the other correlators this
nearest neighbour peak exists but it is usually shifted to a slightly
different value than $q_{max}$. 
This shift can easily be understood. Whereas
for $S_{0 0}(q)$ an average over all possible orientations of the
molecule is done the factor ${\cal D}_{0 0}^l(\Omega_j) {\cal D}_{0
0}^{l'}(\Omega_{j'})$ in eq.(\ref{eq:smolqt}) for $l$ and/or $l'$ unequal
zero "restricts" the average
to a certain range of orientations. Since in the supercooled liquid
the molecules are very closely packed the characteristic distance
between molecules depends on the choice of orientations $\Omega_j$ and
$\Omega_{j'}$ simply because of steric hindrance. Thus for different
correlators a shift in the position of the peaks is expected. A strong
support for this picture is also that the peak is much less shifted for
$S_{0 l'}(q)$ as for $S_{l l'}(q)$ with $l$ and $l'$ unequal zero 
because an average over all possible {\em relative}
orientations is already done if {\em one} of the angles $\Omega_j$ or
$\Omega_{j'}$ is not restricted.

Besides the shift in the peak position we observe that the peak
amplitude also depends strongly on the correlator. One reason already
pointed out by K\"ammerer et al. \cite{kammerer} is the approximate
head--tail--symmetry of the molecule which results in a smaller
amplitude for correlators with $l+l'$ odd. Further we can state that
the amplitude of the main peak tends to decrease with increasing
$l,l'$. 

Apart from the main maximum the center of mass correlation function
$S_{0 0}(q)$ shows also structure at higher values of $q$. Clearly
perceptible is a double peak at $q \approx 10.5$ and $q \approx 13.2$.
Whereas the peak at $q \approx 13.2$ can be identified as the second
order of the nearest neighbour peak the origin of the peak at $q
\approx 10.5$ is not quite clear. It could also be the second order of
a peak located around $q \approx 5.25$ which is merged with the main
peak. 

Rich structure at higher values of $q$ is also found in all other
molecular correlation functions. If one takes into account possible
shifts in the peak position the peaks located at
$q \approx 12-14, q \approx 16-19$ and $q \approx 23-25$ can be
attributed to the higher orders of the nearest neighbour peak. While
their principle origin thus seems to be clear the interesting
structure that appears in their shape, which in some cases clearly
indicates a double peak, can not be understood from the present
investigation.

A further aspect which immediately strikes the eye is that, while the
oscillations at higher $q$ are strongly damped for small $l$ and
$l'$, the higher order peaks are comparable or even larger than the
first maximum for the correlators with $l,l' \in \{2,3,4\}$. 

So far we have only considered the structure of the molecular
correlation functions for $q$--values larger than the value of the
first maximum corresponding to distances smaller than the average
nearest neighbour distance. The center of mass correlator $S_{0
0}(q)$ for $q \le 6.6$ resembles the structure also found for simple
liquids
which is in strong contrast to the behaviour of the molecular
correlation functions for $l,l' \not= 0$. The 
correlator $S_{1 1}(q)$ exhibits a pronounced maximum at
$q \approx 2.8$, a $q$--value corresponding to about twice the average
nearest neighbour distance. 

A further point worth mentioning is that all diagonal correlators $S_{l
l}(q)$ with $l \ge 2$ show a maximum at $q = 0$ as was already found
for  $S_{2 2}(q)$ by K\"ammerer et al. \cite{kammerer}. This might
indicate a tendency for a {\em local} nematic order.

In conclusion of this section we want to point out that a comparison
of our findings with an analysis of the system in real space would be
very valuable for a better understanding of the molecular correlation
functions. It would be especially interesting to connect the shifts in
peak positions, the relation between the peak amplitudes of different
correlators and the peak shapes to a microscopic characterization of
the system in real space.

\subsection{Neutron Scattering Function}
\label{structure factor}

As already mentioned above the results of section \ref{theory} are
best tested for the static case. The molecular correlation functions
necessary for an evaluation of the series representation
eq.(\ref{eq:snresult1}) have been presented in the previous
subsection for $l,l' \le l_{co} = 4$. 
The result of this calculations will be compared to the
exact data for neutron scattering as determined according to
eq.(\ref{eq:site-site}).

Since we are dealing with the special case of linear molecules and
static correlations eqs.(\ref{eq:snresult1})--(\ref{eq:bcoh}) can be
further simplified. As justified in the previous subsection the
indices $n,n'$ can be set to $0$ for the coherent terms if the
$z$--axis of the body--fixed frame of reference is chosen in direction
of the symmetry axis connecting both atoms of the molecule. The polar
coordinates of the atoms A and B are given by 
$(r_A,\vartheta_A,\varphi_A) =
(d/2,0,0)$ and $(r_B,\vartheta_B,\varphi_B) = (d/2,\pi,0)$ respectively.
This information can be inserted into eq.(\ref{eq:bcoh}) for the
coherent molecular form factors. Taking into account
the trivial expression
$S_{ln,l'n'}^{(s) 0}(q) = \delta_{l l'} \delta_{n n'}$ for the static
self correlations and using sum rules for the spherical harmonics and
spherical Bessel functions \cite{abramowitz72} all summations over the
incoherent terms can be carried out leaving only one $q$--dependent
function $b^{inc}(q)$ usually termed molecular structure factor. We
can combine these reformulations to get the following result for the
ns--static structure factor of diatomic rigid molecules.

\begin{eqnarray}
& & S_n(q) \cong S_n^{(l_{co})}(q) = \sum_{l,l' \le l_{co}} b_{l l'}^{coh}(q) 
S_{l l'}^{(d)}(q) + b^{inc}(q) 
\label{eq:snresult2} \\
& & b_{l l'}^{coh}(q) = [ (2l+1)(2l'+1) ]^\frac{1}{2} j_l(\frac{qd}{2})
j_{l'}(\frac{qd}{2}) \; \cdot \label{eq:bcoh2} \\
& & \cdot \; \frac{1}{2} \left( \overline{a_{coh}^A} + (-1)^l
\overline{a_{coh}^B} \right) \left( \overline{a_{coh}^A} + (-1)^{l'}
\overline{a_{coh}^B} \right) \nonumber \\
& & b^{inc}(q) = \frac{1}{2} \left( \overline{(a_{coh}^A)^2} +
\overline{(a_{coh}^B)^2} + \overline{(a_{inc}^A)^2} +
\overline{(a_{inc}^B)^2} \right) + \label{eq:binc2} \\
& & + \overline{a_{coh}^A} \overline{a_{coh}^B} j_0(qd) \nonumber
\end{eqnarray}

We just want to mention that choosing $l_{co} = 0$ yields the usual
Sears-expression \cite{sears67}. This fact demonstrates again that the
representation through molecular correlation functions is a natural
extension of the approach by Sears.

\begin{figure}[htb]
\epsfxsize=8cm
\epsfysize=6cm
\centerline{\epsffile{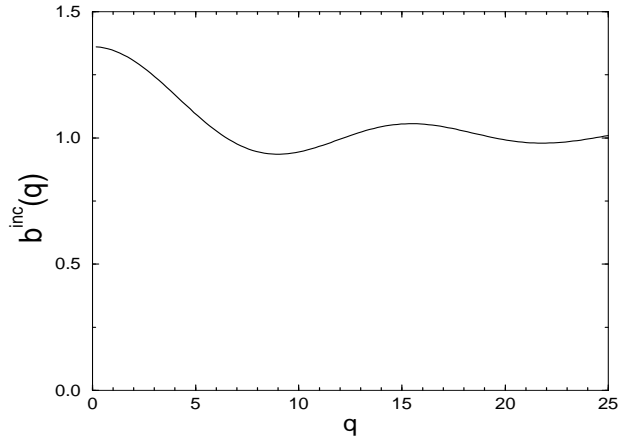}}
\caption{Molecular structure factor $b^{inc}(q)$ for a system of
diatomic rigid molecules 
with $a^A_{coh}=1.4$,$\; a^B_{coh}=0.25$ and $a^\nu_{inc}=0$ for $\nu=A,B$.}
\label{fig:binc}
\end{figure}

The molecular structure factor $b^{inc}(q)$ shown in Fig.
\ref{fig:binc} for the system of diatomic molecules with a special
choice of scattering lengths has a very simple
$q$--dependence. Starting from a maximum at $q = 0$ it quickly decays
and shows oscillations around the asymptotic value for $q \to \infty$
which is given by the
$q$--independent first term of eq.(\ref{eq:binc2}). In the case of
general molecules the $q$--dependence is given by a linear
superposition of functions $j_0(qr_{\nu \nu'})$ involving all
intramolecular distances $r_{\nu \nu'} = | \vec{r}_{j \nu} -
\vec{r}_{j \nu'} |$. 

Both, the molecular correlation functions $S_{l l'}(q)$ and 
the coherent form
factors $b_{l l'}^{coh}(q)$ are symmetric with respect to $l$ and
$l'$. Thus the summation in eq.(\ref{eq:snresult2}) can be restricted
to $l' \ge l$.
If the scattering lengths of A-- and B--atoms are equal the coherent
molecular form factors $b_{l l'}^{coh}(q)$ vanish unless $l$ and $l'$
are both even, i.e. only molecular correlation functions with $l$ and
$l'$ even contribute to $S_n(q)$. This is a specific property of the
diatomic linear molecule closely related to head--tail symmetry. In
case of exact head--tail symmetry the correlators $S_{l l'}^{(d)}(q)$
are zero for $l$ or $l'$ odd. If only the scattering lengths are equal
such correlations may exist but they do not enter $S_n(q)$ since the
neutron probe can not distinguish between A-- and B--atoms. 

\begin{figure}[htb]
\epsfxsize=8cm
\epsfysize=6cm
\centerline{\epsffile{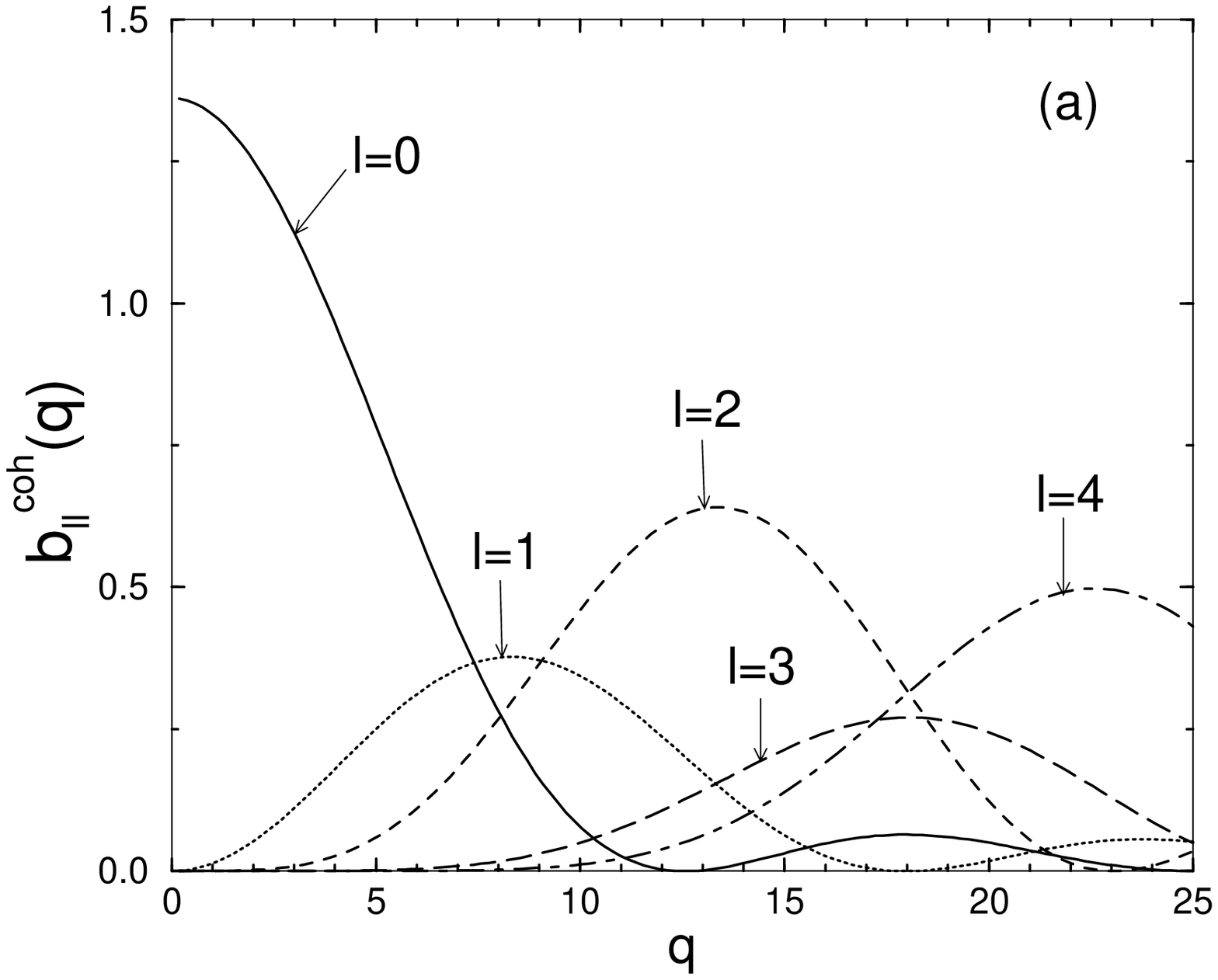}}
\epsfxsize=8cm
\epsfysize=6cm
\centerline{\epsffile{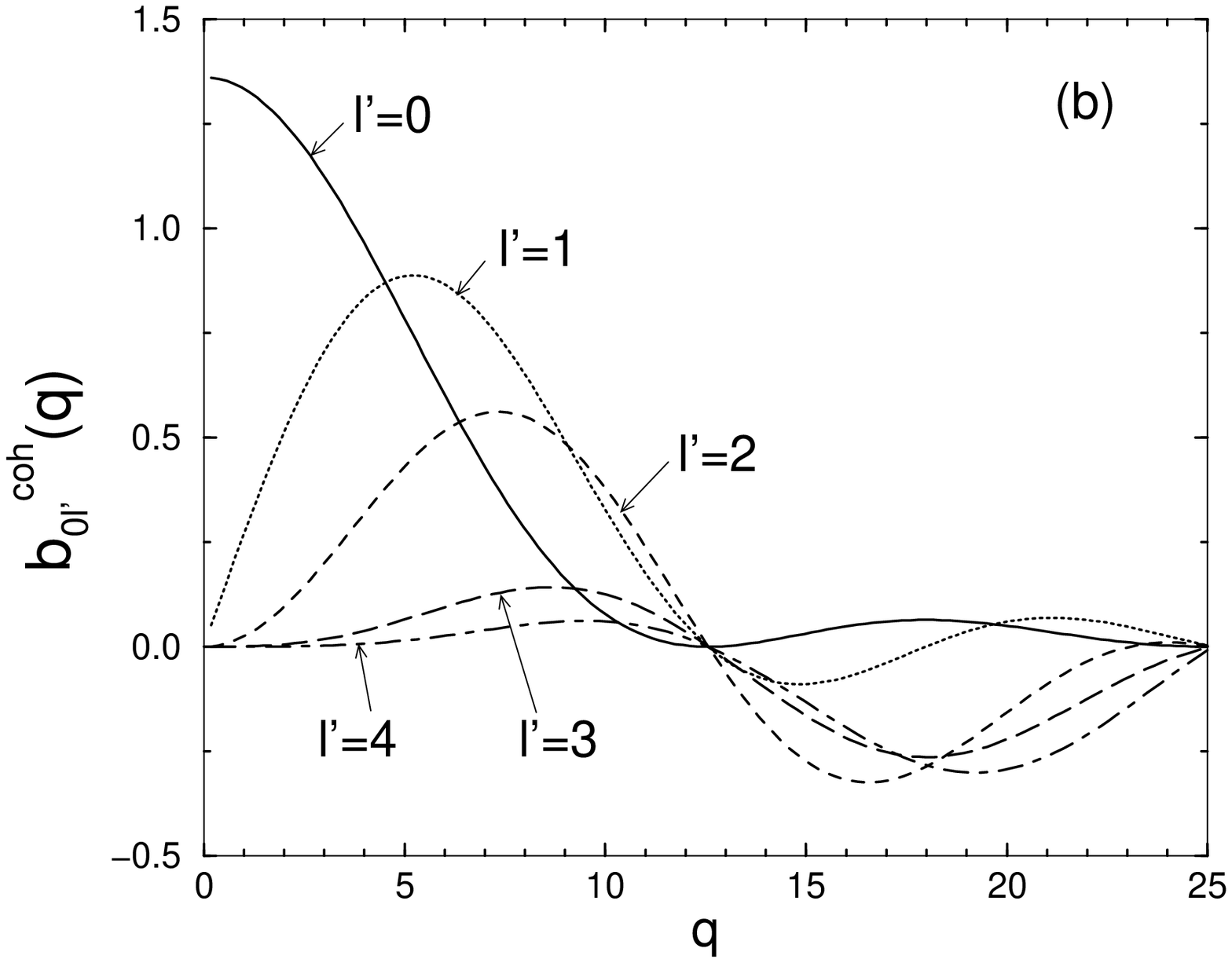}}
\caption{Molecular form factors $b_{l l'}^{coh}(q)$ 
for a system of diatomic rigid
molecules with $a^A_{coh}=1.4$,$\; a^B_{coh}=0.25$ and $a^\nu_{inc}=0$
for $\nu=A,B$. Subfigure (a) shows the progressing shift of the first
maximum to higher $q$ for the diagonal form factors $b_{l
l}^{coh}(q)$, subfigure (b) shows the shift and decrease of the first
maximum for the form factors $b_{0 l'}^{coh}(q)$ with $l = 0$.}
\label{fig:bcoh}
\end{figure}

Apart from
this system--specific property eq.(\ref{eq:bcoh2}) and the examples of
form factors presented in Fig. \ref{fig:bcoh} clarify some properties
of the $q$--dependence of $b_{l l'}^{coh}(q)$ which are quite
universal, i.e. valid also in the time--dependent case and for general
molecules. In both subfigures one can observe that the weight $b_{l
l'}^{coh}(q)$ of a
given correlation function $S_{l l'}(q)$ depends very strongly on the
value of $q$. For instance the center of mass correlations $S_{0
0}(q)$ are very important for $q \approx 0$ but only of minor
relevance for $q > 10$. From the mathematical properties of the
spherical Bessel functions, determining the $q$--dependence of $b_{l
l'}^{coh}(q)$ also in the general case given by eq.(\ref{eq:bcoh}),
one can conclude a systematic relation between the values of $l$ and
$l'$ and the range of $q$--values in which the correlator $S_{l
l'}(q)$ will have a large weight. This relations are demonstrated in
Fig. \ref{fig:bcoh}. Figure (a) shows the diagonal form factors
$b_{l l}^{coh}(q)$ up to $l = 4$. All of them are positive over the
entire $q$--range. $b_{0 0}^{coh}(q)$ has a maximum for $q = 0$ which
decays to zero with increasing $q$ and is followed by further
oscillations with strongly reduced amplitude. The other diagonal
correlators also show a pronounced maximum that is followed by smaller
oscillations but the location of this main maximum is shifted to higher
values of $q$. This shift grows monotonously with increasing $l$.
Further we see that the maximum's amplitude decreases and its width 
gets larger upon increasing $l$. We have chosen $a^A_{coh} =
1.4$ different from $a_{coh}^B=0.25$ in order to have nonvanishing
form factors for $l$ or $l'$ odd. Still we can observe that the
dependence of $b^{coh}_{l l'}(q)$ on the scattering lengths leads to
smaller amplitudes for odd $l$. Fig. \ref{fig:bcoh} (b)
also shows the systematic behaviour for the offdiagonal form factors
$b^{coh}_{0 l'}(q)$ with $l=0$. Upon growing $l'-l$ the amplitude of
the first maximum quickly reduces and the contribution at the second
maximum gets more important. Thus while the offdiagonal terms can not
be neglected the relevant range is also shifted to higher values of
$q$ for growing $l'$. The same is true for the nondiagonal terms with
$l > 0$.

Consequently the systematic behaviour of the
weights $b_{l l'}^{coh}(q)$ can be formulated as a rule of thumb: {\em
The larger the values of $l$ and $l'-l$ the higher the value of $q$ at
which the form factor $b_{l l'}^{coh}(q)$ will become relevant.} We
merely note that for the form factors as well as for the molecular
correlation functions themselves negative values are possible for the
offdiagonal terms.

Having discussed all the relevant terms for the evaluation of the
right hand side of eq.(\ref{eq:snresult2}) we can now turn to the
ns--static structure factor itself and test the convergence of the
series representation, i.e. the quality of the different
approximations $S^{(l_{co})}_n(q)$, $l_{co}=0,1,...,4$.

\begin{figure}[htb]
\epsfxsize=8cm
\epsfysize=6cm
\centerline{\epsffile{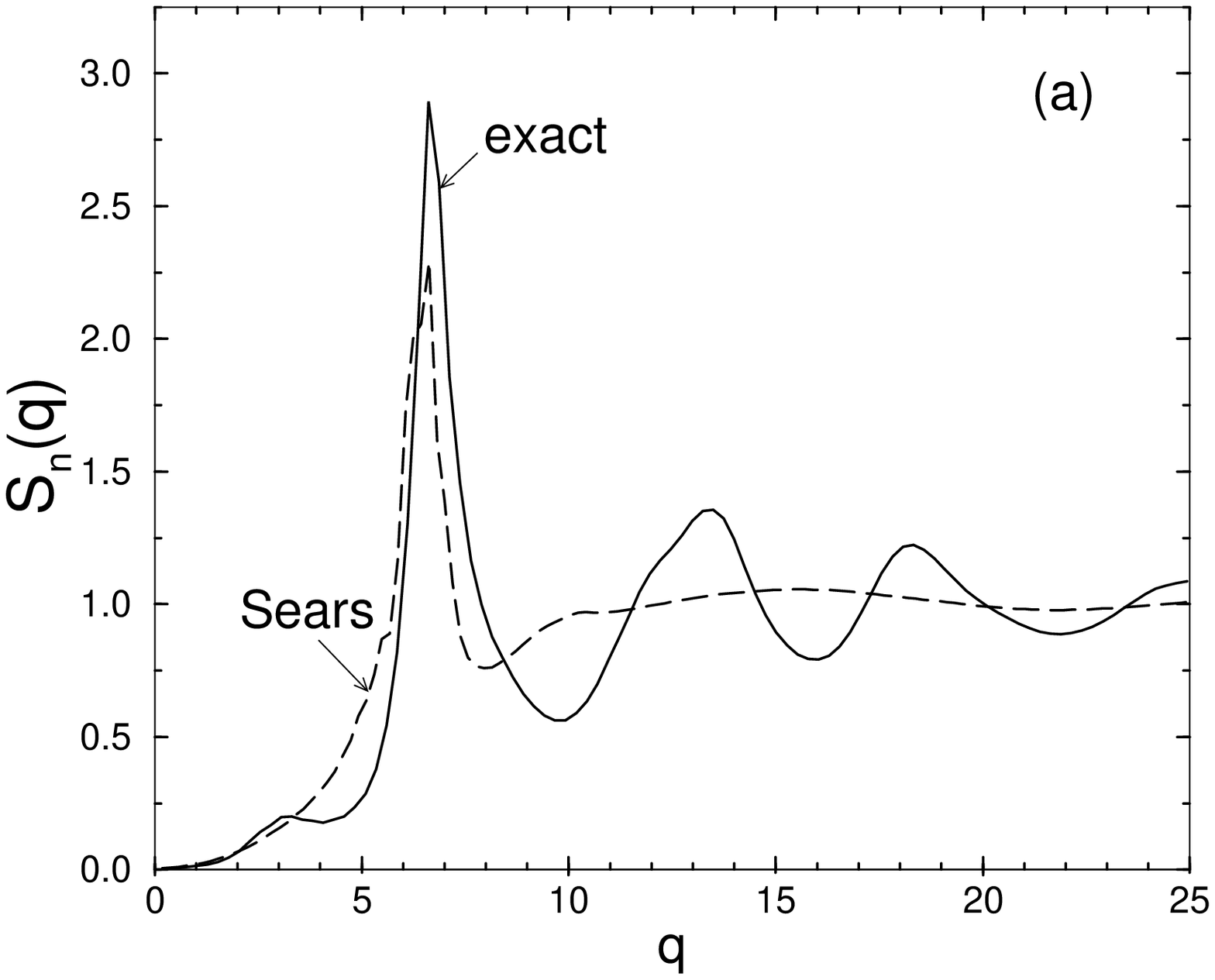}}
\epsfxsize=8cm
\epsfysize=6cm
\centerline{\epsffile{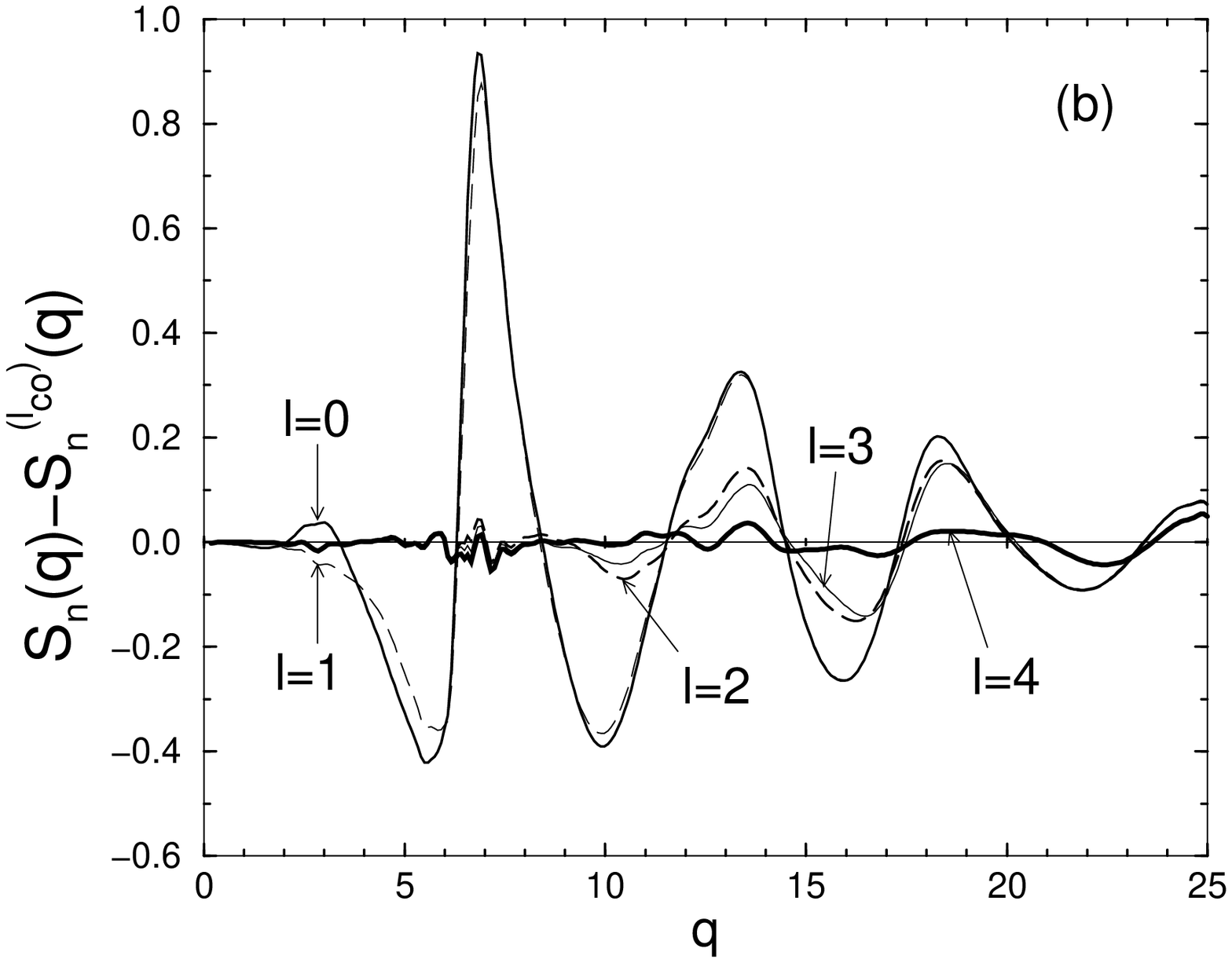}}
\caption{Static structure factor of neutron scattering and its series
representation according to eq.(\protect\ref{eq:snresult2}) for
different values of the $l$--cutoff. The scattering lengths are chosen
as
$a^A_{coh}=1.4$,$\; a^B_{coh}=0.25$ and $a^\nu_{inc}=0$ for $\nu=A,B$. 
(a) exact and Sears--result ($l_{co}=0$). (b) absolute error 
of the approximation for different values of $l=l_{co}$.} 

\label{fig:comp}
\end{figure}

Fig. \ref{fig:comp} (a) shows the exact result for
$S_n(q)$ as evaluated according to eq.(\ref{eq:site-site}). The
scattering length have been chosen as $a^A_{coh}=1.4$,
$a^B_{coh}=0.25$ and $a_{inc}^\nu=0$, $\nu = A,B$. At $q \approx 3$
the static structure factor exhibits a small but well pronounced
prepeak followed by a strong maximum at $q \approx 6.6$ and further
maxima at $q \approx 13$, $q \approx 18$ and $q \approx 25$. Thus the
$q$--dependence of $S_n(q)$ shows the general structure also found for
the molecular correlation functions. Taking a closer look to the
various maxima and taking into account the properties we found for the
weights $b_{l l'}^{coh}(q)$ we can further illuminate their origin. In
case of the prepeak this is rather obvious since $i$) at $q \approx 3$ only
correlators with small $l$ and $l'$ have to be taken into account 
(see discussion above) and
$ii$) only $S_{1 1}(q)$ possesses a significant peak at $q \approx 3$.
Therefore its origin must be
connected to the structure found in the
correlation function $S_{1 1}(q)$. A quantitative analysis of this
statement will be given in the following subsection. From the Figure
\ref{fig:bcoh} we can see that also at the position $q \approx 6.6$ of
the main peak correlators with $l$ or $l'$ larger than 2 do not play
a big role. Thus the center of mass correlation function $S_{0 0}(q)$
and the molecular correlation function $S_{0 2}(q)$,$S_{2 2}(q)$ as
well as $S_{0 1}(q)$ can be identified as the origin of the main peak
in $S_n(q)$. The latter correlators being responsible for the slight
shift of the peak position to higher $q$ in comparison with the center
of mass correlations. For the peaks at $q \approx 13$ and $q \approx
18$ a number of correlators will be relevant. Using the weights $b_{l
l'}^{coh}(q)$
one can still conclude that $S_{2 2}(q)$ will be one of the main
sources for the peak at $q \approx 13$ and $S_{2 4}(q)$ as well as
$S_{4 4}(q)$ will be of great relevance for the structure at $q
\approx 18$. From the molecular correlation functions we have
determined $S_{4 4}(q)$ is the one which contributes most to the peak
at $q \approx 25$ though correlation functions with $l,l' > 4$ could
also be of great importance.

Apart from the exact result for $S_n(q)$ Figure \ref{fig:comp} (a)
also shows the result of the weak hindering approximation as used by
Sears \cite{sears67}. Except for the limiting values for $q \to 0$ and
$q \to \infty$ the results of this approximation are rather poor since
almost all of the structure in $S_n(q)$ is missed and even the main peak is not
reproduced properly. This finding strongly supports results of Chen et al.
\cite{chen97} who observed that a factorization of translational and
rotational correlations is not
suitable for the description of a supercooled liquid.

Figure \ref{fig:comp} (b) shows the convergence 
of the series representation by
presenting the dependence of the error $S_n(q) - S_n^{(l_{co})}(q)$ 
on $l_{co}$. The large error of the Sears result corresponding to
$l_{co}=0$ is already reduced to statistical fluctuations in a range of
$q$ up to $q \approx 9$ by choosing $l_{co}=2$. 
For $l_{co}=4$ the static structure factor $S_n(q)$ is almost
perfectly reproduced for the entire range of $q$--values at least up to
$q \approx 20$. Since the $q$--range chosen here compares well to the values
accessible in a real neutron scattering experiment we can conclude that the
convergence of the series representation (\ref{eq:snresult1}) is fast enough to
make it attractive for practical purposes. The cutoff of the infinite sum at a
relatively small value of $l_{co}$ is well justified, at least for our
choice of model. The actual value
of $l_{co}$ that is sufficient will depend on the size and form of the
molecule as well as their interactions \cite{sciortinopc}.
In accordance with the given rule of
thumb increasing the value of $l_{co}$ merely corresponds to an enlargement of
the range in which $S_n(q)$ is reproduced. 

\subsection{Prepeak}
\label{prepeak}

We will now turn to a special feature of the static structure factor. As already
mentioned in the last subsection $S_n(q)$ shows a prepeak at $q \approx 3$.
Given that prepeaks or first sharp diffraction peaks (FSDP) have been studied
mostly in connection with strong glasses like e.g. $SiO_2$
\cite{elliott91} or ionic glassformers \cite{wilson94} this observation is
quite surprising. In the system of diatomic rigid molecules studied here we
neither find a network structure nor do we have long range electrostatic
interactions. Since the proposed explanations for the prepeak usually rely
heavily on the network structure and/or the effects of the ionic charges they
can not be applied to the prepeak in the present system. Then what could be the
mechanism leading to intermediate range order in a system without an extended
network or long range interactions?

\begin{figure}[htb]
\epsfxsize=8cm
\epsfysize=6cm
\centerline{\epsffile{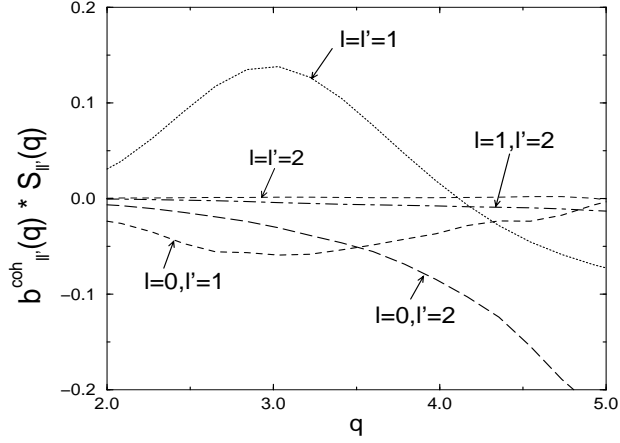}}
\caption{Contributions of the different molecular correlation 
functions to the prepeak for $a^A_{coh}=1.4$,$\; a^B_{coh}=0.25$ and
$a^\nu_{inc}=0$ for $\nu=A,B$.}
\label{fig:prepeak}
\end{figure}

In the preceding sections we have pointed out the connection between the
$q$--dependence of $S_n(q)$ and the contribution of various molecular
correlation functions. To understand the mechanism responsible for the
formation of a prepeak it will surely 
be helpful to analyze in detail which of 
the relevant molecular correlations contribute 
at $q_{pp} \approx 3$. The contributions of
the intramolecular structure factor $b^{inc}(q)$ and the center of mass
correlations $S_{0 0}^{(d)}(q)$ are of large amplitude but they contain no
structure at all at $q_{pp}$. Since they have opposite signs they almost cancel
each other. Together they give the Sears--result which does not
exhibit any prepeak
(cf. Fig.\ref{fig:comp} (a)). 
The other relevant contributions at $q_{pp}$ are shown in Fig.
\ref{fig:prepeak}. Thus the importance of the correlator $S_{1 1}(q)$ for the
formation of the prepeak is corroborated by this quantitative analysis. Besides
the negative contribution of $S_{0 1}(q)$ no other term shows any structure at
$q_{pp}$. It may be noted that the negative contributions of $S_{1 1}(q)$ and
$S_{0 2}(q)$ at $q \approx 5$ induce a better separation between the
prepeak and the main peak. In accord with our "rule of thumb" the terms
involving $l,l' > 2$ are even less important than $S_{2 2}(q)$.

The predominant role of $S_{1 1}(q)$ and the fact that no structure is observed
in the center of mass correlation strongly suggest that orientational effects
are responsible for the prepeak in the studied system. Apart from the evidence
given here it seems not too far fetched to consider spatial ordering due to
sterical hindrance as a mechanism for prepeaks in molecular systems as close
packing \cite{moss85} and size effect \cite{iyetomi93} have also been shown to
be of great relevance in other systems. The importance of orientational degrees
of freedom for the structure of $S_n(q)$ for liquid halogens and 
the prepeak in the molecular liquid $CCl_4$ has also been put forward by Misawa
\cite{misawa89a,misawa90}. Whereas in his work orientational correlations
between neighbouring molecules had to be introduced
as an {\em assumption} we have been
able to give {\em direct} evidence for their relevance.

Taking into account the interpretation of the molecular correlation functions
(see above) it is also possible to get additional information on
the kind of intermediate range order. 
${\cal D}^1_{0 0} \propto Y^0_1$ has the shape
of a dumbbell which is, due to the choice of the body fixed frame, aligned to
the symmetry axis connecting the atoms of the molecule. Thus the correlations
of $S_{1 1}(q)$ at $q_{pp}$ can be attributed to a preference for a parallel
orientation of {\em next} nearest neighbours. Apart from the
geometric effects leading to a parallel alignment of molecules also
energetic affects are important for this "antiferromagnetic" order
\begin{equation}
... \begin{array}{ccc}A & B & A \\ | & | & | \\ B & A & B \end{array}
... , 
\end{equation}
since the vicinity of $A$ and $B$
molecules is favoured by the choice of Lennard--Jones energies
$\epsilon_{AB} < \epsilon_{AA} = \epsilon_{BB}$.

Prepeaks have often been reported to show special behaviour. In many glassy
systems an increase of the prepeak amplitude with increasing temperature is
found \cite{elliott91,salmon94} which is quite contrary to the behaviour of the
other peaks in $S_n(q)$. The amplitude and position was also shown to be
affected by the pressure \cite{elliott91} and the composition of the liquid
\cite{wilson94}. Therefore it will be interesting to characterize further the
behaviour of the prepeak in our system.

\begin{figure}[htb]
\epsfxsize=8cm
\epsfysize=6cm
\centerline{\epsffile{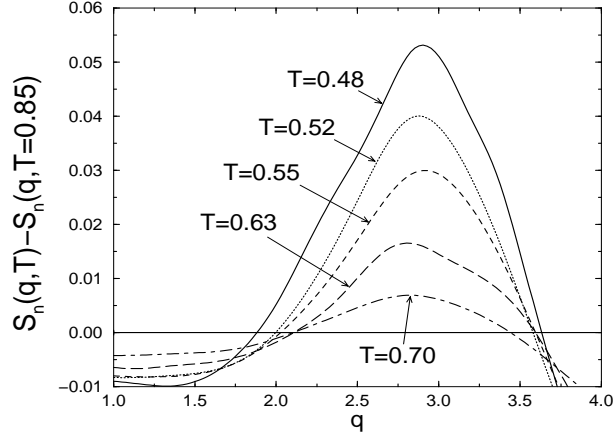}}
\caption{Temperature dependence of the prepeak for
$a^A_{coh}=1.4$,$\; a^B_{coh}=0.25$ and $a^\nu_{inc}=0$ for $\nu=A,B$.}
\label{fig:tdep}
\end{figure}

Fig. \ref{fig:tdep} shows the temperature dependence of the prepeak in the
range $T=0.48$ to $T=0.70$. To eliminate the background given by the low $q$
wing of the main peak we have substracted the $S_n(q)$ at $T=0.85$. At
this temperature the prepeak is not discernible anymore. With increasing
temperature the prepeak decreases in amplitude and the width at half maximum
grows. In the given temperature range no shift in the position of the peak can
be observed whereas the main peak is shifted downwards by $\Delta q=0.2$. 
Although seemingly uncommon
for a prepeak such temperature dependence has been reported before. It has been
found for carbon tetrachloride \cite{misawa89b} and also in a recent study of
m--toluol and m--fluoranilin by Morineau et al. \cite{morineau98}. Like our
system these are molecular liquids. Thus one is tempted to infer that this
behaviour might be typical for such systems.

\begin{figure}[htb]
\epsfxsize=8cm
\epsfysize=6cm
\centerline{\epsffile{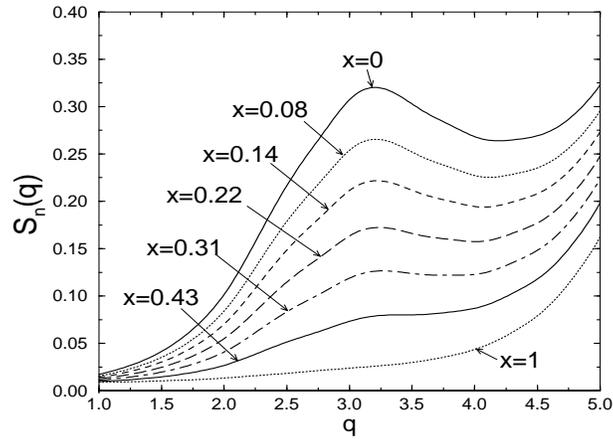}}
\caption{Dependence of the prepeak amplitude on the choice of the
scattering length. $x = a_{coh}^B/a_{coh}^A$ denotes the ratio of the
coherent scattering lengths. The incoherent scattering lengths are set
to zero and $a_{coh}^\nu$, $\nu=A,B$ are chosen in such a way that
the large $q$ limit of $S_n(q)$ is constant.}
\label{fig:adep}
\end{figure}

We are not able to determine the pressure dependence of the prepeak 
or to study effects of composition, i.e. the influence of 
atomic size and interactions since this has not been done in the
simulation \cite{kammerer}. 
Still we want to point out one aspect which is closely related to
compositional studies but has never been considered before. A substitution of 
atoms of a certain species in the liquid will not only alter the interactions
and the size of the atoms but will also affect the scattering lengths. Figure
\ref{fig:adep} shows that the observability of the intermediate range order as
a prepeak can indeed be influenced. As can also be inferred from
eq.(\ref{eq:snresult2}) in the case $a_{coh}^A = a_{coh}^B$, i.e. 
$x=a_{coh}^A/a_{coh}^B=1$, the correlator $S_{1 1}(q)$ gives no contribution to
$S_n(q)$ and the prepeak vanishes. Keeping the overall normalization constant
and increasing the asymmetry in the scattering lengths the prepeak grows to its
maximum value. Since at the small $q$--value of 
the prepeak position the molecular
correlation function with $l=1$ will be the most important one (apart from
$S_{0 0}(q)$) a prepeak in molecular systems will be most easily observed for
molecules with {\em no} head--tail--symmetry. Thus it is for instance not
surprising that in liquid halogens no prepeak is observed although their
liquid structure might be not too different from our system \cite{misawa89a}.

\section{Conclusion}
\label{conclusion}

Tackling the problem of neutron scattering in supercooled molecular
liquids we have shown that well known concepts can be generalized
naturely to give a concise description including all molecular degrees
of freedom. Thus it was possible to reduce the large discrepancy that
was found between the exact result for $S_n(q)$
and a description in terms of
the center of mass only. The resulting description of $S_n(q,t)$ 
in terms of molecular correlation functions $S_{ln,l'n'}^m(q,t)$ was
shown to  shed light on the $q$--dependence offering clear relations
between the $q$--range and the angular resolution described by the
"quantum numbers" $l$ and $n$. Giving a relation between the
intermediate scattering function and the molecular correlations the
representation could be used to extract (partial) information on the
correlations of TDOF as well as ODOF and their interference from
neutron scattering experiments. From the general expression we were
able to give hints for an effort in this direction. A quantitative
test of the formalism for a liquid of diatomic molecules in the "worst
case" of static correlations led to a good agreement with exact
results for $S_n(q)$ if molecular correlation functions up to $l_{co}
= 4$ are taken into account. This analysis also confirmed that a
factorization like the weak hindering approximation which leads to a
description in terms of the center of mass only is not suitable in the
supercooled regime.

In the simulation of a liquid of diatomic Lennard--Jones molecules we
observe a prepeak in the ns--static structure factor $S_n(q)$ which
could be attributed to intermediate range orientational order. The
temperature dependence of this prepeak is in accord with the results
found for the molecular liquids $CCl_4$, m--toluidine and
m--fluoroaniline but in variance with the behaviour found for most
covalent glassformers. The influence of scattering lengths on the
observability of intermediate range (orientational) order was examined
offering the conclusion that a manifestation as a prepeak may not
occur in case of head--tail symmetry of the molecules.

\acknowledgements

We are very grateful to S.K\"ammerer and W.Kob who performed the
simulation work and provided us with the necessary data.
We also thank E.Bartsch for pointing out some of the references given
below. 
This work was supported financially by SFB 262 which is gratefully
acknowledged as well.

\end{multicols}

\onecolumn

\newpage

\newpage

\newpage

\end{document}